\documentstyle[12pt]{article}

\newcommand{\Tr}{{\rm Tr}}
\newcommand{\ie}{{\em i.e.~}}
\newcommand{\eg}{{\em e.g.~}}

\newcommand{\G}{{\Gamma}}

\newcommand{\be}{\begin{equation}}
\newcommand{\ee}{\end{equation}}
\newcommand{\ba}{\begin{eqnarray}}
\newcommand{\ea}{\end{eqnarray}}
\newcommand{\bea}{\begin{eqnarray}}
\newcommand{\eea}{\end{eqnarray}}
\newcommand{\bl}{\Bigg(}
\newcommand{\br}{\Bigg)}

\def\a{\alpha}
\def\b{\beta}
\def\c{\gamma}

\def\e{\epsilon}           
\def\f{\phi}               
  
\def\g{\gamma}

\def\r{\rho}                                     

\def\D{\Delta}
\def\F{\Phi}
\def\G{\Gamma}

\def\cn{{\cal N}}
\def\co{{\cal O}}

\def\half{{1 \over 2}}

\def\bop#1{\setbox0=\hbox{$#1M$}\mkern1.5mu
        \vbox{\hrule height0pt depth.04\ht0
        \hbox{\vrule width.04\ht0 height.9\ht0 \kern.9\ht0
        \vrule width.04\ht0}\hrule height.04\ht0}\mkern1.5mu}
\def\pa{\partial}                              
\def\de{\nabla}                                       

\def\>{\rangle} 

\def\<{\langle} 
\def\Dsl{D \hskip-.6em \raise1pt\hbox{$ / $ } }

\def\to{\rightarrow}
\def\tf{\tilde{\f}}
\def\pa{\partial}

\def\+{\oplus}

\def\de{\mbox{d}}

\def\nonu{\nonumber \\{}}
\def\half{{1 \over 2}}
\def\Tr{{\rm Tr}\, }

\begin{document}
\begin{flushright}
 ROM2F/2003/02
\\
hep-th/0302xxx
\end{flushright}
\begin{center}
{\large\bf Holographic three-point functions:} 
\\
{\large \bf one step beyond the tradition}
\vskip .8truecm

{Massimo Bianchi\footnote{{\tt Massimo.Bianchi@roma2.infn.it}} and
Alessandro Marchetti\footnote{{\tt crius@tiscali.it}}}

$^1${\it Dipartimento di Fisica and INFN \\
Universit\`a di Roma ``Tor Vergata''\\
00133 Rome ITALY}\\
$^2${\it Dipartimento di Fisica and INFN \\
Universit\`a di Roma ``La Sapienza''\\
00185 Rome ITALY}

\vskip 0.5truecm

\end{center}
\vskip .5truecm
\begin{abstract}
Within the program of holographic renormalization, we 
discuss the computation of three-point correlation functions along RG 
flows. We illustrate the procedure in two simple cases.
In an RG flow to the Coulomb branch of ${\cal N}=4$ SYM theory
we derive a compact and finite expression for the three-point function of 
lowest CPO's dual to inert scalars. 
In the GPPZ flow, that captures some features of  ${\cal N}=1$ SYM
theory, we compute the three-point function with insertion
of two inert scalars and one
active scalar that mixes with the stress tensor. 
By amputating the external legs at the mass poles 
we extract the trilinear coupling of 
the corresponding superglueballs.
 Finally we outline the procedure for computing
three-point functions with insertions of the stress tensor as well as
of (broken) R-symmetry currents.
\end{abstract}

\section{Introduction}
Holographic renormalization \cite{bfs1,bfs2} is a powerful tool that 
produces finite correlation functions and allows explicit checks of 
(anomalous) Ward identities along RG flows dual to
asymptotically AdS domain walls. One of the main achievement of the
program is an efficient and consistent algorithm for computing finite
one-point functions in the presence of external sources. By sources we mean the
coefficient functions, two per each bulk field component, which are not 
fixed by the near-boundary analysis \cite{hs,dhss}. One of those corresponds 
to an operator deformation of the fixed point theory, while the other 
corresponds to giving a VEV to an operator \cite{kw}.

The non-local and non-linear relation between VEV and
source which amounts to requiring regularity of the
classical solution of the non-polynomial bulk field equations is
hard to find in general. Luckily in order to compute two-point correlation
functions it is enough to know this relation at the linear level,
\eg by solving the linearized fluctuation equations around a given
domain wall solution. Disentangling the mixing between
longitudinal/trace modes of the stress tensor and the active
scalar(s) proved subtler \cite{bfs1,wm},
than initially expected \cite{dwf,aft}. Similar difficulties 
appearing in the mixing of the broken symmetry currents
with Goldstone bosons were disposed of in \cite{bfs2,bs2,mm}. 
Transverse modes are easier to analyze since
their fluctuations can be expressed in terms of an auxiliary
massless scalar \cite{bdwfp,bs1}. In particular, after getting rid of
non-canonical kinetic terms, (transverse) vector fields display a universal
effective mass $M^2=-2A''$ \cite{bdwfp,kwy,kras} that 
is manifestly positive thanks to the
convessity of the scale factor $A(r)$ which is at the heart of the holographic 
c-theorem \cite{fgpw1,gppz1}. Interesting mass spectra emerge from  
momentum analysis of holographic two-point functions but so far  
tradition has prevented the study of even the simplest interactions.
 
There is little hope of computing higher-point correlation functions 
except possibly for three-point functions\footnote{Following the procedure
of holographic renormalization, three-point 
functions at the conformal point dual to pure AdS 
have been computed by D.~Freedman, U.~G\"ursoy and K.~Skenderis as reported 
in \cite{sken}.}. Higher point functions
require the knowledge of the bulk-to-bulk propagator that appears
in exchange diagrams. For three-point functions, the relevant
Witten diagrams at tree-level only involve bulk-to-boundary
propagators that are known in many cases, \eg solving the
linearized fluctuation equations with prescribed boundary
conditions. Bulk-to-bulk propagators would only appear in loop
diagrams that are suppressed by the 5-D Newton constant \ie by
inverse powers of $N$ in the spirit of the AdS/CFT correspondence
\cite{jm,edw,gkp}\footnote{For recent reviews see 
\eg \cite{magoo,dhf,mb,sken} and references therein.}. 
Thus restricting to the planar limit
captured, at strong 't Hooft coupling, by the supergravity approximation
one can go a step beyond the 
tradition 
and address the problem of computing three-point functions. 
By `computing' we actually mean writing down closed-form expressions
that allow one to extract interesting information about the interactions 
in particular regimes. The only ingredient that is
missing from previous analyses is the exact form of some bulk
cubic couplings along the flows and of some additional counterterms. 
In this note we fill this gap 
and compute simple three-point functions of both active and inert scalars 
in the two most studied holographic RG flows: the GPPZ flow
\cite{gppz2} and the CB flow \cite{fgpw2,bs1}. 

In order to fix the notation we briefly review the procedure of holographic 
renormalization in Sect. \ref{holoren}. We then describe the
general strategy for computing three-point functions 
in Sect. \ref{strate} and collect some basic
formulae for the two flows 
in Sect. \ref{rgflows}. The only real novelty that deserves a special 
mention is the identification
of additional logarithmically divergent counterterms 
in the GPPZ flow when the bulk field dual to the gluino bilinear comes into 
play.

In Sect. \ref{cbflow} we specialize our analysis to the CB flow. 
With little effort we write down a compact expression for 
the three-point
function of inert scalars belonging to the $({\bf 3}, {\bf 3})$
representation of the $SO(4)$ subgroup of $SO(6)$ preserved along the  
flow. We check that our result is finite for generic momenta and try to 
convince the reader that it 
displays the correct UV behavior for large momenta. For comparison and
in order to give a flavor of the complexity of the problem, 
we compute the Fourier transform of the conformal
three-point functions of scalar primary operators which
is not easy to find in the existing literature \cite{tod}. 
We thus report the relevant 
formula for $\Delta =2$ in an appendix.

In Sect. \ref{gppz} we analyze the three-point
function with one insertion of the active scalar that interpolates for the
trace of the stress tensor and two insertions of the inert scalars, which is 
a ``caricature'' of the gluino bilinear in the GPPZ flow. 
As expected the resolution of the mixing with the stress tensor 
at the required order in the 
fluctuations turns out to be quite laborious but 
straightforward. The naive result is logarithmically divergent in 
the contact terms that are cancelled by the additional 
counterterms previously identified. 
By amputating the external legs we extract the finite and rather explicit
trilinear coupling of the relevant superglueball states. 
 
Ideally one would like to compute three-point correlation functions
with insertions of the stress tensor and/or of (broken) R-symmetry currents. We
plan to return to this problem soon. In the concluding Section 
we outline the procedure for computing such
three-point functions and speculate how similar computations,
following or extending when needed the approach of holographic
renormalization along the lines of \cite{kras},
should be feasible for theories which
are dual to full-fledged string solutions possibly with logarithmic
corrections even in the UV \cite{mn,ks,kt}. 

\section{Holographic Renormalization strikes back}
\label{holoren}

In this section we fix our notation. Since there only minor changes with 
respect to \cite{bfs1,bfs2,sken}, readers familiar with holographic 
renormalization may 
skip directly to Section \ref{strate}.

To be specific we focus on asymptotically AdS domain walls which are
dual to RG flows in $d=4$. 
The set of fields driving the flow together with some `spectator' fields
represents a consistent truncation of some gauged supergravity in
$D=5$, which in turn is the low-energy effective theory of some
superstring compactifications on an Einstein manifold with
stabilizing fluxes. As usual, we
consider Poincar\'e invariant solutions of the form
\cite{fgpw1,fgpw2,gppz1,gppz2,bs1}
\be 
ds^2 = e^{2A(r)} dx\cdot dx + dr^2 \qquad \Phi = \Phi(r)
\label{ansatz}
\ee 
where $\Phi$ denote one or more canonical scalar fields that drive the flow.
The AdS boundary (field theory UV) is at $r\to +\infty$. The (IR)
interior presents a naked singularity unless the flow terminates
at another (super)conformal fixed point.
Preserving some supersymmetry reduces the problem to solving first order
`gradient flow' equations \be A'(r) = - {g\over 3} W(\Phi)
\qquad {\Phi}'(r) = {g\over 2} \pa_{\Phi}W \ee 
where $g=2/L$ is the bulk gauge coupling and $L$ 
is the characteristic AdS scale.
The scalar potential reads
\be V = {g^2\over 4}\left[{1\over 2} 
\left({\pa W\over \pa\Phi}\right)^2 - {2\over 3} W^2\right] 
\label{potential}\ee

To extract some physics from the domain wall solution
one may not forgo a near-boundary analysis of
the non-linear field equations \cite{dhss}. To this end it is customary
to introduce a different radial variable $z$ such that $dr = - L
dz/z$, \ie $z = \exp(-r/L)$. The scale factor behaves as $A(r)
\approx r/L\approx -\log z$ near the UV boundary at
$z=0$. When $\exp(2A) \approx \exp(+2r/\tilde{L})$ as
$r\to -\infty$ one has an AdS `horizon' with $\tilde{L}< L$ that
describes a different fixed point at the endpoint of the flow
\cite{fgpw2,bs}.
Barring string corrections or uplift to $D=10$ \cite{pw1,pw2}, many interesting
solutions display a naked singularity at some finite value of the
radial variable.
Depending on the behavior of the active scalar field(s) near the
boundary one has flows along which (super)conformal invariance is
either spontaneously broken, by the presence of operator VEV's, or
explicitly broken, by a relevant deformation of the fixed point
action.

For scalar fields dual to operators with UV dimension $\Delta$,
\ie with AdS mass $(ML)^2=\Delta(\Delta -4)$, the near boundary
expansion, that reads 
\be 
\Phi(z, x) = 
z^{4-\Delta} (\phi_{(0)}(x) + z \phi_{(1)}(x) + ...) +
z^{\Delta}(\tilde\phi_{(0)}(x) + z \tilde\phi_{(1)}(x)+ ...)
\ee 
displays two {\it a priori}
independent asymptotic behaviors. The iterative solution of the
field equations allows one to express the higher order coefficient
functions $\phi_{(n)}$ and $\tilde\phi_{(n)}$ in terms of
$\phi_{(0)}$ and $\tilde\phi_{(0)}$.  For $\Delta$ an integer,
only coefficient
functions with `even' indices appear and it is convenient to change
radial variable to $\r = z^2$. Moreover the two dominant behaviors
differ by an integer and additional logarithmic terms are needed
in order to satisfy the field equations\footnote{In the very special case
$\Delta = 2$, even the dominant term contains a logarithm $
\Phi(\r,x) = \r \log\r \phi_{(0)}(x) + \r\tilde\phi_{(0)}(x) + ...$.}. 
Logarithmic behaviors
can be associated to conformal anomalies that, being local, can be
written in terms of $\phi_{(0)}$, playing the role of an
operator source, and a finite number of derivatives thereof. It is
not difficult to see that $\tilde\phi_{(0)}$ is related to the
VEV of the corresponding operator \cite{kw,dhss}. 

Naive extension of the AdS/CFT correspondence to holographic RG flows 
suggests one to compute 
the on-shell action as a functional of the boundary 
data and then take functional derivatives in order to compute correlation 
functions. However the unregulated on-shell action is infinite due to the
infinite volume of the bulk. A convenient regularization consists
in a radial cutoff $\rho>\epsilon$ that allows one to easily 
isolate a finite number of terms diverging as $\e \rightarrow 0$. 
For the case of scalars coupled to gravity the on-shell
action takes the schematic form \be \label{reg} S_{{\rm reg}}[\phi_{(0)},
g_{(0)};\e]
=  \int_{\r=\e} \de^{4}x \sqrt{g_{(0)}} \left[ {a_{(0)}\over \e^{\nu}}
+ {a_{(2)}\over \e^{\nu-1}} 
+...-\log \e \ a_{(2 \nu)} + \co(\e^0)\right]
\nonumber \ee where $\nu = 2$
and $a_{(2k)}$ are local functions of the
sources, $\phi_{(0)}$ and $g_{(0)}$, that appear in the expansions of $\Phi$ 
and of the metric
\be
g_{(0)ij}= g_{(0)ij} + \r g_{(2)ij} + \r^2 g_{(4)ij} + \r^2 \log\r h_{(4)ij}
+ ...
\label{metrexp}
\ee

The counterterm action $S_{{\rm ct}}[\Phi, \gamma;\e]$, chosen in
such a way as to cancel the divergent terms in $S_{{\rm
reg}}[\phi_{(0)},g_{(0)};\e]$, is to be
expressed in terms of the fields $\Phi(x,\e)$ `living' at the
regulating surface $\r=\e$ and of the induced metric, $\c_{ij} =
g_{ij}(x,\e)/\e$. This is required for covariance and follows from
a straightforward but often laborious  ``inversion'' of the
expansions (\ref{metrexp}),(\ref{ansatz}). 
The counterterms are universal in that they
make the on-shell action finite for {\it any} solution of the
bulk field equations with given boundary data. The counterterms
are however different for different (consistent) 
truncations, \eg for different potentials $V(\F)$.
So far we have described a ``minimal'' scheme in which only the
divergences of $S_{\rm reg}$ are subtracted. As in standard
quantum field theory, one has the freedom to add finite
invariant counterterms. These correspond to a change of scheme aimed
for instance to restoring some (super)symmetry.

For later purposes, it proves convenient to  
define a subtracted action at the cutoff \be \label{renact} S_{{\rm
sub}}[\Phi,\gamma;\e] = S_{{\rm reg}}[\phi_{(0)},g_{(0)};\e] + S_{{\rm
ct}}[\Phi,\gamma;\e]. \ee 
The subtracted action has a finite limit
as $\e \to 0$, and the renormalized action is a functional of the
sources defined by this limit, \ie \be \label{sren} S_{\rm
ren}[\phi_{(0)},g_{(0)}] = \lim_{\e \rightarrow 0} S_{{\rm
sub}}[\Phi,\gamma;\e] \ee The distinction between $S_{\rm sub}$ and
$S_{\rm ren}$ is needed because the variations required to obtain
correlation functions are performed before the limit $\e
\rightarrow 0$ is taken.
In particular, the expectation value of a scalar operator in the presence of 
sources, defined by \be
\label{1ptdef} \< \co_\phi \> = {1 \over \sqrt{g_{(0)}}} {\delta
S_{\rm{ren}} \over \delta \f_{(0)}} \ee 
where $g_{(0)}= \det(g_{(0)ij})$ can be computed by
rewriting it in terms of the fields living at the regulating
surface\footnote{For scalars of UV dimension $\Delta=2$, an additional
$\log\e$ is needed in this formula.} \be
\label{oexp} \< \co_\phi\> = \lim_{\e \to 0} \left( {1 \over
\e^{\D/2}} {1 \over \sqrt{\g}} {\delta S_{\rm{sub}} \over \delta
\F(x,\e)} \right) \ee where $\g = \det(\g_{ij})$. In
general one can prove that \be \< \co_\phi\> = (2\Delta - 4)
\tf_{(0)} + {\rm local} \ee where the local terms are completely
fixed by the choice of $\f_{(0)}$.

The hard problem is to determine the non-local relation between
the operator VEV $\tf_{(0)}$ and the operator source $ \f_{(0)}$
that guarantees regularity of the solution in the deep interior
(IR regime) where the supergravity solution may develop a
singularity. Traditionally 
only two-point function have been explicitly computed along
holographic RG flows \cite{bfs1,bfs2,bs1,bs2,wm,mm,bs}. 
These only require the non-local relation at
linear order \be \tf_{(0)}(x) = {1\over 2\Delta - 4} \int d^4x'
C_2(x-x') \f_{(0)}(x') \ee or better, in momentum space, \be
\tf_{(0)}(p) = {1\over 2\Delta - 4} C_2(p) \f_{(0)}(p) \ee that is
established by solving the linearized fluctuation equations with
prescribed boundary conditions, \ie regularity in the interior.
Upon differentiation one gets the two-point correlation function
\be \< \co_{\f}(x) \co_{\f}(x') \> = C_2(x-x') + {\rm contact
\ terms} \ee

\section{Three-point functions: general strategy}
\label{strate}

Our aim in this section is to sketch the further 
necessary steps that one should take in 
order to compute three-point functions.
For simplicity we detail the procedure for
inert scalar fields with cubic couplings, a case that finds 
application in Section \ref{cbflow}. 
Mixing with
metric fluctuations or with vector fields is not discussed
in this section. We will address this issue in Section \ref{gppz}.

For three-point functions one needs to take into account quadratic
terms in the field equations. To this end consider the Euclidean
action for a real 
inert scalar around a domain wall\footnote{The overall constant is determined 
by careful reduction from $D=10$ \cite{gkp}.}  
\be S = {N^2\over 2\pi^2} \int d^4x
d\r \sqrt{G} [{1\over 2} Z(\r)(\pa\Phi)^2 + {1\over 2} M^2(\r)\Phi^2
+ {1\over 3} T(\r)\Phi^3 + ...] \ee 
The effective
mass $M(\r)$ and the cubic coupling $T(\r)$ may depend on the
radial variable $\r$ via their functional dependence on the active
scalar(s). The kinetic term $Z(\r)$ may also depend on $\r$. In the
cases we explicitly discuss here scalar fields are canonical. In
order to keep  our formulae as compact as possible we thus assume
$Z=1$ henceforth. 

The non-linear field equation is of the form
\be (\nabla^2_G - M^2) \Phi = T \Phi^2 + ... \ee

At the classical level the solution can be found by summing
trees. Loops are suppressed by powers of $1/N^2$ and we 
neglect them since they can only be consistently taken into
account after embedding the model in string theory. 
For a given boundary source $j(x)$ we find
\bea
&&\Phi(x,\r) = \int d^4x' K(x,\r;x') j(x') \ + \\
&& \int d\r' d^4x'\sqrt{G} D(x,\r;x',\r') T(\r')
\left[\int d^4x'' K(x',\r';x'') j(x'')\right]^2 + ...\nonumber \eea
where $D(x,\r;x',\r')$ is the bulk-to-bulk propagator,
$K(x,\r;x')$ is the bulk-to-boundary propagator and $...$ denote
higher-order terms in $j(x)$.

The linearized approximation, which is enough to compute two-point
functions, yields \be \Phi_{(1)}(\r, p) = K(\r, p) j(p) \ee in
momentum space. The near boundary anlysis then 
gives\footnote{The second index denotes the order in $j$.}
\be
\phi_{(0|1)}(p) = j(p) \ee and 
\be \tilde\phi_{(0|1)}(p) = {1\over 2\Delta - 4} C_2(p)
j(p) \ee
To quadratic order in $j$ \be \Phi_{(2)}(\r, p) = \int
d\r' \sqrt{G} D(p,\r;\r') T(\r') \int d^4q K(q,\r')
j(q) K(p-q,\r') j(p-q) \ee
Expanding both sides in powers of $\r$ yields
 \be \phi_{(0|2)}(p) = 0 \ee and \be
\tilde\phi_{(0|2)}(p) ={1 \over 2\Delta - 4} \int 
d\r \sqrt{G} K(p,\r) T(\r) \int d^4q K(q,\r) j(q) K(q-p,\r) j(p-q) \ee
since near the AdS boundary, where $\r\approx 0$,
\be D(x,\r;x',\r') \approx
{\r^{\Delta/2} \over 2\Delta - 4} K(x;x',\r') \ee

Plugging this back into the on-shell action or better into the
one-point function given above, differentiating w.r.t. 
$j$
and then setting $j$ to zero, one finally gets \be \langle
\co_\phi(p_1) \co_\phi(p_2)\co_\phi(p_3) \rangle = (2\pi)^4
\delta(\sum_i p_i) \int_{\r>\epsilon} d\r \sqrt{G(\r)} T(\r) \prod_i
K(p_i,\r) \ee 
modulo contact terms that are polynomial in the momenta and cancel potential 
divergences arising when the regulator of the radial integration is removed. 

In all cases we consider, the linearized fluctuation
equations reduce to hypergeometric equations in a suitable radial 
variable $w=w(\r)$. The solution
which is regular in the deep IR interior (which is a naked
singularity of the `good' kind \cite{gubs}) is of the form \be
\phi_p(w,x) = w^\a(1-w)^\b F(a,b;c;w)\exp(ipx) \ee
Analytically continuing the hypergeometric function from $w=0$,
that represents the deep IR interior (singularity),  to $w=1$,
that represents the UV boundary, determines the
bulk to boundary propagators. The reader may find some 
useful formulae in appendix B.

\section{Our favorite holographic RG flows}
\label{rgflows}

For completeness we gather in this section some relevant formulae
about two particularly simple yet interesting 
asymptotically AdS domain wall solutions: the CB flow \cite{fgpw2,bs1}
and the GPPZ flow \cite{gppz2}.
The former is dual to an 
RG flow from ${\cal N} =4$ SYM at its maximally superconformal
point to a locus in the Coulomb branch  where
superconformal symmetry is broken spontaneously. The latter is dual to 
an RG flow to a 'confining' ${\cal N}=1$ SYM theory where superconformal
invariance is broken explicitly by turning on relevant (mass)
operators for the three chiral multiplets. 

\subsection{Coulomb branch flow} \label{cbdisc}

The first case we consider is the CB flow with $n=2$, in the
notation of \cite{fgpw2}. The supersymmetric domain wall solution 
describes a continuous distribution of D3-branes on a disk.
The relevant superpotential  is given
by \be W(\F)=-e^{-{2 \F /\sqrt{6}}} - \half e^{{4 \F /
\sqrt{6}}}. \ee Using (\ref{potential}), one can easily compute
the potential. Near $\F=0$ it can be expanded as 
\be \label{CBpot}
V(\F) = -{3\over L^2} - {2\over L^2} \F^2 
+ {4 \over 3 \sqrt{6}L^2} \F^3 + \co(\F^4) \ee
which exhibits a tachyonic mass $M^2L^2 = -4$, as appropriate for
a field dual
to a chiral primary operator (CPO) of UV dimension $\D =2$.

It is very convenient to express the domain-wall solution in terms of a
new radial variable $v$. The solution and the relation between $v$ and $r$
are \be   
\F = {1\over \sqrt{6}}\log v, 
\qquad 
e^{2A}={\ell^2\over L^2} {v^{2/3} \over
1-v}, \qquad {d v \over d r} = {2\over L} v^{2/3} (1-v) 
\nonu \ee 
The
boundary is at $v=1$ and the solution has a curvature singularity 
at $v=0$. The parameter $\ell$ is the radius of the disk of branes
\cite{fgpw2,bs} in the 10-dimensional ``lift'' of this solution and sets the 
scale of a mass gap in the spectrum of excitations to $m_{o} = \ell/L^2$.

The scalar field vanishes at the boundary at the rate \be \F
\approx -{1 \over \sqrt6} (1-v) = - {1 \over \sqrt6} \exp(-2r/L) \ee
that corresponds to the dual operator of UV dimension $\D =2$ taking a VEV.
The endpoint of the flow describes $\cn = 4$ SYM 
theory at a locus in
the Coulomb branch where the $SO(6)$ R-symmetry is broken
to $SO(2) \times SO(4)$. For any finite $N$,
the $SO(2)$ symmetry is actually 
broken \cite{bfs2,bs2}. The scalar that gets a VEV
is the neutral  singlet component of the CPO
$Q^{ij}_{\bf 20'}= Tr~(X^{i} X^{j})\vert_{\bf 20'}$ that decomposes according 
to
${\bf 20}' \rightarrow ({\bf 1}, {\bf 1})_0 \+ ({\bf
1}, {\bf 1})_{+2} \+ ({\bf 1}, {\bf 1})_{-2} \+ ({\bf 2}, {\bf
2})_{+1} \+ ({\bf 2}, {\bf 2})_{-1} \+ ({\bf 3}, {\bf 3})_0 $
under $SO(4) \times SO(2)$. 

\subsection{GPPZ flow}

The GPPZ flow \cite{gppz2} corresponds to adding an
operator of dimension $\Delta = 3$ to the Lagrangian, 
namely the top component of the
superpotential $\Delta {\cal W} = \sum_{I} Z_I^2$, that gives a common 
mass to the three ${\cal N} =1$  chiral multiplets $Z_I$ appearing
in the decomposition of the ${\cal N}=4$ vector multiplet. The
solution was proposed as the holographic dual of pure ${\cal N}=
1$ SYM  theory \cite{gppz2}. Although it does not capture all of the expected
properties of the field theory, it is particularly simple and
still displays some interesting features including a discrete spectrum of 
superglueballs.

The active scalar is a singlet under an $SO(3)$ subgroup of
$SO(6)$. A consistent truncation to the $SO(3)$ singlets yields
${\cal N}=2$ gauged supergravity coupled to two hypermultiplets
describing a $G_{2(2)}/SO(4)$ coset \cite{pw1,bdwfp}. 
After lengthy calculation one
gets the $5d$ superpotential that reads \be W(\F, \Sigma)=-{3 \over 4}
\left[\cosh \left({2 \F \over \sqrt{3}}\right) + \cosh \left({2
\Sigma }\right)\right] = -{3 \over 2} -{1 \over 2}(\F^2 +
\Sigma^2) + ... \ee 
Near $\F=\Sigma=0$, the potential 
admits the expansion 
\be \label{gppzpot} V(\F, \Sigma) = - {3\over L^2}
-{3 \over 2L^2} (\F^2 + \Sigma^2) - {1 \over 3L^2} ( \F^4 - 3 \Sigma^4 +
6 \F^2 \Sigma^2) + ... \ee 
The mass of both $\Phi$ and $\Sigma$
is $M^2L^2 =-3$, indicating that the dual scalar operators have
UV dimension $\Delta = 3$. For simplicity we only consider the 
domain-wall solution with $\Sigma = 0$
\footnote{More general solutions with a non-trivial profile
for the bulk field $\Sigma$ dual to the gaugino bilinear have been
considered by GPPZ.}, and \be \F = {\sqrt{3}\over 2} \log {1
+\sqrt{1-u} \over 1-\sqrt{1-u}} \qquad e^{2 A} = {u \over 1-u},
\ee where $u = 1 - \exp(-2r/L)$. The boundary is at $u=1$ and the
solution has a naked singularity of `good' type at $u=0$. Since
$\F \approx \sqrt{3} \exp(-r/L)$ near the boundary, we are dealing
with an operator deformation rather than a
VEV. Along the flow \bea &&W(u) = -{3 \over 2u} \qquad W_\phi(u) =
-\sqrt{3}{\sqrt{1-u} \over u}
\\
&&W_{\phi\phi}(u) = - {2-u \over u} \qquad W_{\phi\phi\phi}(u) =
-{4\over \sqrt{3}}{\sqrt{1-u} \over u} \eea

For our later purposes, we need the effective mass of the field 
$\Sigma$ along the GPPZ flow
\be M^2_\sigma(u) = V_{\sigma\sigma} = -{3\over L^2}\left[2\cosh
\left({2 \F \over \sqrt{3}}\right) -1 \right] = 
-{3\over L^2} \ {4-3u \over u}
\ee 
and its cubic coupling to the active scalar $\Phi$ 
\be\pa_\phi M^2_{\sigma}(u)= V_{\phi\sigma\sigma}= -
{4\sqrt{3}\over L^2} \ \sinh\left({2 \F \over \sqrt{3}}\right)=
-{8\sqrt{3}\over L^2} \ {\sqrt{1-u} \over u} \ee
also along the flow.

\section{Three-point functions in CB flow}
\label{cbflow}

In order to simplify our lives we only consider
inert scalars of UV dimension $\Delta = 2$,
\ie those among the 20 tachyonic scalars which are not
singlet under the $SO(4)\times SO(2)$ preserved in the CB flow. In
particular we concentrate our attention on the $({\bf 3},{\bf 3})_0$
components  $Q^{i_L i_R}$ and aim to compute
\be
{\cal G}_{CB}^{i_L i_R,j_L j_R,k_L k_R}(x_1,x_2,x_3) =
\< Q^{i_L i_R}(x_1) Q^{j_L j_R}(x_2) Q^{k_L k_R}(x_3)\>
\ee
The rationale behind this choice is that there is a 
unique $SO(4)\times SO(2)$ invariant trilinear coupling among the $Q$'s. 
It is proportional to
$\varepsilon_{i_Lj_Lk_L} \varepsilon_{i_Rj_Rk_R}$. 
In order to avoid additional derivative couplings, it is necessary to
identify three components of  $Q^{i_L i_R}$ with 
canonical kinetic term around the CB domain wall solution.
It is well known that the
icosaplet parametrizes a coset $SL(6)/SO(6)$. Taking as coset
representative a symmetric matrix ${\cal S}$ with 
$\det({\cal S})=1$, the scalar
potential reads \be V = - {g^2\over 32} [ (tr({\cal S}))^2 - 2 tr({\cal S}^2)]
\ee ${\cal S}$ can be diagonalized by an orthogonal $SO(6)$
transformation \be {\cal S} = R(\theta) \tilde{{\cal S}}(\beta) 
R^T(\theta) \ee where 
$\tilde{S}(\beta)
= diag(\exp(2\beta_1), ..., \exp(2\beta_6))$ with $\sum_i \beta_i =
0$. The six
$\beta$'s should then be expressed in terms of five
independent fields. A convenient choice that leads to
canonically normalized scalars 
is 
\bea 
&&\beta_1 = {\sqrt{2}\over\sqrt{3}} [-\Phi + \sqrt{3} (\lambda +\sigma +
\omega)]
\nonumber \\
&&\beta_2 = {\sqrt{2}\over\sqrt{3}} [-\Phi + \sqrt{3} (\lambda -\sigma
-\omega)]
\nonumber \\
&&\beta_3 = {\sqrt{2}\over\sqrt{3}} [-\Phi + \sqrt{3}(-\lambda +\sigma
-\omega)]
\nonumber \\
&&\beta_4 = {\sqrt{2}\over\sqrt{3}} [-\Phi + \sqrt{3} (-\lambda -\sigma
+\omega)]
\nonumber \\
&&\beta_5 = {\sqrt{2}\over\sqrt{3}} [ 2 \Phi + \sqrt{6}\nu]
\nonumber \\
&&\beta_6 = {\sqrt{2}\over\sqrt{3}} [ 2 \Phi - \sqrt{6}\nu] 
\eea 
Taking $\Phi$ as active scalar, one can convince oneself
that $\nu$ transforms under the lower $SO(2)$ subgroup while
$\lambda,\sigma,\omega$ only transform under the upper $SO(4)$. 
The former is part of the charged singlet $({\bf 1},{\bf 1})_\pm$. 
The latter three are part of
the neutral ennaplet $Q^{i_L i_R}\in({\bf 3},{\bf 3})_0 $.

We can proceed by switching off all the other inert scalars, setting $\Phi$ 
to its background value
and keeping only terms up to cubic order in
$\lambda, \sigma,\omega$. We
find \be V(\Phi, \lambda, \sigma,\omega, ... ) = 
V(\Phi) + {1\over 2}M^2(\Phi) [\lambda^2 +
\sigma^2 + \omega^2] + T(\Phi)
\lambda\sigma\omega + ... \ee 
As expected $\lambda, \sigma,\omega$ have the same
effective mass \be M^2(\Phi) = -
{4\over L^2}e^{+2 \Phi/\sqrt{6}} \ee and a unique
trilinear coupling \be T(\Phi) = {1\over \sqrt{2} L^2}[
2 e^{-4 \Phi/\sqrt{6}}- e^{2\Phi/\sqrt{6}}] \ee
Along the CB flow \be M^2(v) = -
{4\over L^2} v^{1/3} \qquad
 T(v)= {1\over \sqrt{2} L^2} v^{-2/3}(2-v)
\ee The relevant wave equation for the inert scalars  
$Q(x,v)  = e^{ipx} Q(v)$ reads \be Q'' + \left( {2\over v} + {1\over
1-v}\right) Q' + \left[{p^2\over 4 \ell^2}{1\over v^2(1-v)} +
{1\over v(1-v)^2}\right]Q = 0 \ee 
with $Q=\lambda,\sigma,\omega$.
The solution that vanishes at
the singularity $v=0$ is \be Q(v) = v^a (1-v)
F(a,a;2a;v)\ee where\footnote{Using 
$P^{(0)}_\nu (z) = F(-\nu, \nu +1; 1; (1-z)/2)$, 
$F(a,a;2a;v)$ can be expressed in terms of associated Legendre functions,
but little is gained in this rewriting.}
\be 
a(p) = -{1\over 2} + {1\over 2}\sqrt{ 1 + {p^2\over m_o^2}} \ee

The inclusion of the fields $\lambda,\sigma$ and $\omega$ does not modify in 
any significant way the near boundary analysis performed in \cite{bfs1,bfs2}.
In particular the renormalized one-point function of the operator dual
to any one of them (say $\lambda$) is simply given by
\be
\< \co_\lambda (x) \> = 2 \tilde\lambda_{(0)}(x)
\label{onelambda}
\ee

Analytic continuation of $F(a,a;2a;v)$ 
reveals the non-local relation between $\tilde\lambda_{(0)}$
and $\lambda_{(0)}$ at the linearized level and yields the two-point function 
\be
\< \co_\lambda (p) \co_\lambda (q) \> = 4 {N^2\over 2\pi^2} (2\pi)^4 
\delta(p+q) [\psi(1) - \psi(a)]
\ee
where the correct normalization of the action has been taken into account.
The spectrum is continuous above a mass gap $m_{o}=\ell/L^2$.
Similar analysis leads to the bulk-to-boundary propagator (in momentum space)
\be K_p(v) = - {\Gamma(a)^2 \over \Gamma(2a)}  v^a (1-v) F(a,a;2a;v) \ee 

Taking the mixed derivative of (\ref{onelambda})
w.r.t. the sources of the remaining two fields
($\sigma_{(0)}$ and $\omega_{(0)}$) gives the desired three-point function  \be
\langle \co_\lambda(p_1) \co_\sigma(p_2)\co_\omega(p_3) \rangle =
(2\pi)^4 \delta(\sum_i p_i) \int_0^1 dv \sqrt{G(v)} T(v) \prod_i
K_{p_i} (v) \ee Plugging in the relevant expressions yields 
\be
{\cal G}_{_{CB}}(p_1,p_2,p_3) = \kappa_{_{CB}} \delta(\sum_i p_i) 
\prod_i {\Gamma(a_i)^2 \over \Gamma(2a_i)}
\int_0^1 dv (2-v) \prod_i v^{a_i} F(a_i,a_i;2a_i;v) 
\ee
where $ \kappa_{_{CB}} = - (2 N^2/\sqrt{3})(2\pi)^2(\ell^4/L^2)$.
Since the $SO(4)=SU(2)_L\times SU(2)_R$
tensor structure is unique, Wigner-Eckart theorem gives
\be
\< Q^{i_L i_R}(p_1)Q^{j_L j_R}(p_2)Q^{k_L k_R}(p_3) \> =
\varepsilon^{i_L j_L k_L}\varepsilon^{i_R j_R k_R}{\cal G}_{_{CB}}(p_1,p_2,p_3)
\ee

Although the integral cannot be performed elementarily, it is easy to see that 
${\cal G}_{_{CB}}(p_1,p_2,p_3)$ 
is finite for generic momenta, as expected for operators of UV dimension
$\Delta =2$, and vanishes at the threshold 
$p_i^2 = - m_o^2$ where $a_i = -1/2$.
Series integration yields a compact expression
\be
{\cal G}_{_{CB}}(p_1,p_2,p_3) = \kappa_{_{CB}}  
 \sum_{n_i} \prod_i {\Gamma(a_i+n_i)^2 \over \Gamma(2a_i+n_i)n_i!} 
\left[ {2\over 1 + \sum_i (a_i + n_i)} - {1\over 2 + \sum_i (a_i + n_i)}\right]
\ee
where a $\delta(\sum_i p_i)$ is understood. 

Direct comparison with the conformal three-point function of scalar primary 
operators of dimension $\Delta = 2$ computed in appendix A seems hopeless.
However using the AdS/CFT correspondence for pure AdS we know that
\be
{\cal G}_{_{CFT}}(x_1, x_2, x_3)= 
\kappa_{_{CFT}} T(0)\int {dz d^4x \over z^5} 
\prod_i {z^2 \over [z^2 + (x-x_i)^2]^2} 
\ee
can be written in momentum space, up to constants, as
\be
{\cal G}_{_{CFT}}(p_1, p_2, p_3) = \tilde{\kappa}_{_{CFT}}\delta(\sum_i p_i)
\int {dz \over z^5} T(0) \prod_i z^2 {\cal K}_0(zp_i) 
\ee
where 
\be
{\cal K}_0 (u)= \sum_{n=0}^{\infty} {u^{2n}\over 2^{2n}(n!)^2}
[\log(u/2) - \psi(n+1)]
\ee
is a modified Bessel function.
Changing radial variable according to\footnote{The exact coordinate 
transformation with $\r=z^2/L^2$ can be found in 
\cite{bfs2}.}
\be
1-v = {\ell^2\over L^4}z^2 - {2\over 3} {\ell^4\over L^8}z^4 + \co (z^6)
\ee
it is easy to convince oneself that the UV limit $\ell\to 0$ `explodes' the 
boundary region $v\approx 1$ where indeed the asymptotics 
of $K(p,v)$ coincides 
with the expansion of $z^2{\cal K}_0(zp_i)$ and the trilinear coupling goes to 
its fixed point value $T(0)$. This fixes the UV behavior including the 
overall normalization. 

\section{Three-point function in the GPPZ flow}
\label{gppz}

Although the exact form of the fully non-polynomial potential  
of the ${\cal N}=2$ gauged supergravity that governs the GPPZ flow
is known\footnote{We thank G. Dall'Agata for disclosing its secret form 
to us.}, it is not easy to isolate trilinear couplings that only 
involve inert scalars. Partly for this reason and partly to push our analysis 
another step forward, we are lead to consider 
the still relatively simple 3-point function
\be
{\cal G}_{_{GPPZ}}(x_1,x_2, x_3) 
= \< \co_\sigma(x_1) \co_\sigma(x_2)\co_\phi (x_3)\>
\ee
where $\co_\phi$ is the operator dual to the active scalar $\Phi$ and 
 $\co_\sigma$ is the operator dual to the inert scalar $\Sigma$.
${\cal G}_{_{GPPZ}}$ vanishes at the superconformal fixed point in the UV 
by $SO(6)$ R-symmetry arguments.

The situation is subtler than in the previous Section. On the one hand 
$\Phi$ mixes with the longitudinal and trace components of the metric. 
On the other hand a naive analysis inspired by 
the ``dynamical scalar gauge'' shows 
that the above 3-point function is logarithmically divergent as
$\epsilon\to 0$ if no new counterterm is generated when $\Sigma$ is included. 
However, slightly at
variance to what happens when only $\Phi$ is present, we find two
new logarithmically divergent 
counterterms. One is quartic in $\Sigma$ the other is biquadratic in $\Phi$ 
and $\Sigma$. 

In order to derive the precise form of these counterterms one has to
perform a detailed near boundary analysis of the bulk field equations.
The results we need are 
\bea && g_{ij}^{(2)} = {1\over 2}
(R_{ij}^{(0)} - {1\over 6}R^{(0)}g_{ij}^{(0)})
-{1\over 3}(\phi_{(0)}^2 +\sigma_{(0)}^2) g_{ij}^{(0)} \\
&& h_{(4)} = - 2 (\phi_{(0)}\tilde\phi_{(2)}+ 
\sigma_{(0)}\tilde\sigma_{(2)}) \\
&& g_{(4)} = {1\over 4}(g_{(2)}g_{(2)}) - 2 (\phi_{(0)}\phi_{(2)}+
\sigma_{(0)}\sigma_{(2)}) +4 (\phi_{(0)}\tilde\phi_{(2)}+
\sigma_{(0)}\tilde\sigma_{(2)}) \nonumber \\
&& \quad + {1\over 9} ( \phi_{(0)}^4 - 3 \sigma_{(0)}^4 + 6 \phi_{(0)}^2
\sigma_{(0)}^2)\\
&&\tilde\phi_{(2)}= -{1\over 4} (\nabla_{(0)}^2\phi_{(0)} -
(g_{(2)})
\phi_{(0)}) -{1\over 3} \phi_{(0)}^3 - \phi_{(0)}\sigma_{(0)}^2 \\
&&\tilde\sigma_{(2)}= -{1\over 4} (\nabla_{(0)}^2\sigma_{(0)} -
(g_{(2)})
\sigma_{(0)}) + \sigma_{(0)}^3 - \phi_{(0)}^2\sigma_{(0)} \\
\eea
where $R_{(0)}= g^{ij}_{(0)}R_{(0)ij}$,
$g_{(2)}= g^{ij}_{(0)}g_{(2)ij}$, $g_{(4)}= g^{ij}_{(0)}g_{(4)ij}$,
$ h_{(4)}= g^{ij}_{(0)}h_{(4)ij}$ and 
$(g_{(2)}g_{(2)})= g^{ij}_{(0)}g_{(2)jl}g^{lk}_{(0)}g_{(2)ki}$.

Using Einstein equations in the convenient form
\be
R_{\mu\nu} = -2 (\pa_\mu \Phi \pa_\nu \Phi+ 
\pa_\mu \Sigma \pa_\nu \Sigma + {2\over 3} V
G_{\mu\nu})
\ee
and including the Gibbons-Hawking boundary term, 
it is easy to see that the (regulated) on-shell action is simply
given by \be 
S_{\rm reg} = - {2\over 3}\int_{\rho>\e} d\r d^4x \sqrt{G} V +
\epsilon \pa_\epsilon \int {d^4 x \over \epsilon^2}
\sqrt{g(x,\epsilon)} 
\ee 
Inserting the data arising from the near boundary analysis gives
\bea
&& S_{\rm reg}[g_{(0)}, \phi_{(0)}, \sigma_{(0)}; \e]  = 
\int d^4x \sqrt{g_{(0)}} \left\{
- {3\over 2\epsilon^2} + {1\over 2\epsilon} (\phi^2_{(0)}+ \sigma^2_{(0)})
\right.\\ 
&&\left. + \log\epsilon \left[
 {1\over 32}
\left(R_{(0)ij}R_{(0)}^{ij}- {1\over 3} R_{(0)}^2\right) 
+  {1\over 2} (\phi_{(0)}\nabla^2_{(0)} \phi_{(0)} +
\sigma_{(0)}\nabla^2_{(0)} \sigma_{(0)}) \right.\right.
\nonumber \\
&&\left. \left.+  {1\over 12}
R_{(0)}(\phi^2_{(0)}+ \sigma^2_{(0)}) - 2 \sigma^4_{(0)}
+ 2 \phi^2_{(0)}\sigma^2_{(0)}\right] + \co (\e^0) \right\}
\nonumber
\eea

Once expressed in terms of the regulated
fields, the counterterms read \bea 
&& S_{\rm ct}[\gamma,\Phi,\Sigma;\e] = \int_{\rho=\epsilon} d^4x
\sqrt{\gamma} \left\{ {3\over 2}
- {1\over 8} R[\gamma] + {1\over 2} (\Phi^2 + \Sigma^2) \right.\\
&& - \log\epsilon \left[ {1\over 4}\left(\Phi\nabla_{[\gamma]}\Phi
+ {1\over 6} R[\gamma]\Phi^2\right) + {1\over
4}\left(\Sigma\nabla_{[\gamma]}\Sigma
+ {1\over 6} R[\gamma]\Sigma^2\right) \right.\\
&&\left.\left.+ 2 \Sigma^4 - 2 \Sigma^2\Phi^2 + {1\over 32}
\left(R_{ij}[\gamma]R^{ij}[\gamma]- {1\over 3} R[\gamma]^2\right)
\right]\right\} \eea As anticipated the symmetry between $\Phi$ and $\Sigma$ is
spoiled by the logarithmically divergent quartic terms at the beginning 
of the last line.

The one-point function of ${\cal O}_\phi$ in the presence of
sources is \be \langle {\cal O}_\phi \rangle = \lim_{\e\to 0}
\{ 2 (\phi_{(2)} +
\tilde{\phi}_{(2)}) - {2\over 9} \phi_{(0)}^3 - 4\log\e
\phi_{(0)}\sigma_{(0)}^2 \}\ee
Despite appearance $\< {\cal O}_\phi\>$ is finite as we will 
see. 

In order to compute the desired three-point function we need to
determine the (non-local) dependence of $\tilde{\phi}_{(2)}$ on
$\sigma_{(0)}$. Because of the mixing of the fluctuation  $\varphi$
of the active
scalar $\F$ with the longitudinal ($H$), radial ($h_{rr}$) and
trace ($h=  g^{ij}_{(0)}h_{ij}$) 
components of the metric one has to study the coupled field
equations to linear order in $\varphi$, $h_{rr}$, $h_{ij}$ and $H$
and to quadratic order in $\sigma$. Following the steps detailed by
\cite{dwf,aft}, but keeping $h_{rr}$, one gets \bea &&h' +{16\over 3}
\Phi' \varphi - 4 A'h_{rr} = J_C
\\
&&H'' + 4 A'H' - {1\over 2} e^{-2A} (h + 2 h_{rr}) = J_H
\\
&&3 A' (h' - \pa^2 H) + {3\over 4} e^{-2A}\pa^2 h + 4V h_{rr} - 4
\Phi' \varphi' - 4 V_\phi \varphi = J_G \eea
 where $\pa^2 = \delta^{ij}\pa_i\pa_j$ and 
\bea &&J_C =
{16\over 3\pa^2} \pa^i (\sigma' \pa_i\sigma)= {16\over 3\pa^2}
\pa^i T^{(2\sigma)}_{io}
\\
&&J_H = {4e^{-2A} \over 3\pa^2} \left(4{\pa^i\pa^j\over \pa^2} 
- \delta^{ij}\right)\pa_i\sigma \pa_j\sigma =
{4e^{-2A} \over 3\pa^2}\left(4{\pa^i\pa^j\over \pa^2} 
- \delta^{ij}\right) T^{(2\sigma)}_{ij}
\\
&&J_G = 2 (\sigma')^2 - 2e^{-2A}\pa^i\sigma
\pa_i\sigma - 2M_\sigma^2\sigma^2  = 4 T^{(2\sigma)}_{oo}
\eea 
Introducing the `gauge invariant'
combinations 
\be P= h + {{16 W \over 3W_\phi}} \varphi \qquad R =
h_{rr} + 2 \pa_r \left({\varphi\over W_\phi} \right) \qquad Q = H'
- 2e^{-2A}{\varphi\over W_\phi} \ee 
one gets \bea
&& P' - 4 A' R = - J_C \\
&& Q' + 4 A' Q - {1\over2}e^{-2A} P -  e^{-2A} R = J_H \\
&& 3 A' P' - 3 A' \pa^2 Q + {3\over 4}e^{-2A}\pa^2 P +4 V R = J_G \eea
Manipulating these equations as in \cite{dwf,bfs1}
eventually boils down to
\be 
R'' + (2
W_{\phi\phi} - 4 W)R' + (2 W_{\phi\phi\phi}W_{\phi} - 4 W_{\phi}^2
+ {32\over 9} W^2 -  {8\over 3} W W_{\phi\phi}- e^{-2A}p^2) R = J_R
\ee where the $\sigma$-dependent source 
\be J_R = - 4 \sigma'\sigma' - {4\over 3} M^2 \sigma^2 -
\pa_r \left({\sigma^2 \over W_\phi} {\pa M^2\over \pa_\phi}\right)
- 2 A' \left({\sigma^2 \over W_\phi} {\pa M^2\over
\pa_\phi}\right) \ee 
turns out to be relatively simple thanks to the identity
\be J'_G + 4 A' J_G + {3\over 4}\pa^2 e^{-2A}
J_C -3 A' J_H + 3A' J'_C + 12 (A')^2 J_C = -2 \pa_r (M_\sigma^2) \sigma^2
\ee 
that follows from
\be
\nabla^\mu_B T^{(2\sigma)}_{\mu\nu}= - {1\over 2}\pa_\nu (M_\sigma^2)
\sigma^2 \ee 
where $\nabla^\mu_B$ is the background covariant derivative.

Switching to the standard variable $u$ and to momentum space yields
\be R'' + {1\over u(1-u)} R' + {1\over u(1-u)} \left( {2u-1\over
u(1-u)} - {p^2L^2\over 4}\right) R = J_R(u) \ee 
The solution of
the homogeneous equation is 
\bea 
&&R^{(0)}_p= u(1-u) F(a^{(\phi)}_+, a^{(\phi)}_-; 3;u) = 
{u(1-u)\Gamma(3) \over \Gamma(a^{(\phi)}_+) \Gamma(a^{(\phi)}_-)} 
\sum_{n=0}^\infty {(a^{(\phi)}_+)_n(a^{(\phi)}_-)_n\over (n!)^2}\times
\nonumber\\
&& \times (1-u)^n [2
\psi(n+1) -\psi(n+ a^{(\phi)}_+) -\psi(n+a^{(\phi)}_-) - \log(1-u)] 
\eea 
where
\be
a^{(\phi)}_\pm= {3\over 2} \pm {1\over 2} \sqrt{1 - p^2L^2}
\ee

For $u\approx 1$, $R(u)$ admits an
expansion of the form \be R = R_{(0)}(1-u)\log(1-u) +\tilde{R}_{(0)} (1-u)
+   .. \ee In the axial gauge $h_{rr}=0$ one finds
\be
\tilde{R}_{(0)}  = - {4\over\sqrt{3}}
({\phi}_{(2)}+ \tilde{\phi}_{(2)} - {\phi}_{(0)})
\qquad  {R}_{(0)}= - {4\over\sqrt{3}} \tilde{\phi}_{(2)}
= {1\over\sqrt{3} } \nabla_{(0)}^2\phi_{(0)} + ...
\ee
To linear order around the GPPZ flow, where ${\phi}_{(0)}=\sqrt{3}$ and 
$\tilde{\phi}_{(0)}=1/\sqrt{3}$, while $\tilde{\phi}_{(2)}={\sigma}_{(0)}=0$,
\be 
 \< \co_\f\> = - {\sqrt{3}\over 2} \tilde{R}_{(0)}
\ee

Differentiating $\tilde{R}_{(0)}$
 w.r.t. $\f_{(0)}$, brings in an overall factor of $p^2$ that shows up in 
the two-point function
\be
\langle \co_\phi(p) \co_\phi(-p)\rangle = {N^2\over 4\pi^2}
p^2 \left[ \psi( a^{(\phi)}_+(p)) + \psi( a^{(\phi)}_-(p)) -
2\psi(1)\right]
\ee 

A similar though 
simpler analysis \cite{dwf,aft} leads to the other two-point function
\be
\langle \co_\sigma(p) \co_\sigma(-p)\rangle = {N^2\over 4\pi^2}
\left(p^2 - {8\over L^2}\right) \left[ \psi( a^{(\sigma)}_+(p)) + 
\psi( a^{(\sigma)}_-(p))\right]
\ee 
where
\be
 a^{(\sigma)}_\pm (p) = {3\over 2} \pm {1\over 2} \sqrt{9-p^2L^2}
\ee
Using \be \psi(z) = -\gamma +
\sum_{n=0}^\infty {z-1 \over (n+1) (z + n)} \ee 
it is easy to identify the mass poles \cite{bdwfp} and compute their 
positive (!) residues. 

For $\co_\f$, that belongs to the ${\cal N}= 1$
`anomaly' multiplet ${\cal A} = \sum_I { \Tr} (Z_I^{2})$ 
dual to the active hypermultiplet, one has 
\be
\label{Aspec} (m_\f L)^2 = 4 (k+1)(k+2) \qquad k=0,1,2,\ldots  \,. \ee
For  $\co_\sigma$, that belongs to 
the ${\cal N}= 1$ `Lagrangian' multiplet ${\cal S} = {
\Tr} (W^{2} + ...)$ 
 dual to the dilaton hypermultiplet,
\be 
\label{Sspec} (m_\sigma L)^2 = 4 k(k+3) \qquad k=0,1,2,\ldots \,, \ee
including a zero-mass pole.

It is amusing to observe that the superglueball decay constants arising from
the residues at the $k^{th}$ mass pole are the same for $\phi$ and $\sigma$ 
\be |f_\phi(k)|^2 = {N^2 m^2_\phi(k)
(2k+3)\over \pi^2 L^2}  = |f_\sigma(k)|^2 \ ! 
\ee 

We are ready to compute the three-point function. To quadratic order in 
$\sigma$'s
\be
\< \co_\f (p) \>_\e = - {\sqrt{3}\over 2}
\tilde{R}_{(0)}(p) = - {\sqrt{3}\over 2}
\int_0^{1-\e} du \sqrt{G(u)} K_R(u,p)J_R(u,p)
\label{rego}
\ee
where, setting the active scalar to its background,
\be
J_R(u) = -16 (\sigma')^2 -32 (1-u) \sigma' \sigma -12\sigma^2 
\ee
Differentiating (\ref{rego}) twice w.r.t. $\sigma_{(0)}$ 
one gets the logarithmically divergent integral\bea
&& {\cal G}_{_{GPPZ}}(p_1,p_2, p_3) = \kappa _{_{GPPZ}}
\delta(\sum_i p_i)\prod_i\Gamma(a_+(p_i)) \Gamma(a_-(p_i))
\int_{0}^{1-\epsilon} 
du \ u^3(1-u)^2  \times \nonumber \\
&& F_R (p_3) \left[ (F'_\sigma(p_1)F_\sigma(p_2)+
F_\sigma(p_1)F'_\sigma(p_2) ) - 2
(1-u)F'_\sigma(p_1)F'_\sigma(p_2)\right] \eea 
where $\kappa_{_{GPPZ}} = (N^2\sqrt{3}/2\pi^2 L)(2\pi)^4$ and
\be
 F_R (p)=  F(a^{(\phi)}_+(p), a^{(\phi)}_-(p); 3;u)
\ee
and
\be
 F_\sigma (p)=  F(a^{(\sigma)}_+(p), a^{(\sigma)}_-(p); 3;u)
\ee
The logarithmic divergence of $\tilde{R}_{(0)}$ is a contact term 
independent of the momenta. Its coefficient 
$\eta = 4\sqrt{3}$ is exactly cancelled by the last contact term in 
$\<\co_\phi\>$ when ${\phi}_{(0)}$ is put to its background value of 
$\sqrt{3}$. So that $\<\co_\phi\>$ as well as 
${\cal G}_{_{GPPZ}}$ are finite as they should. 
It is relatively easy to check that ${\cal G}_{_{GPPZ}}$ 
vanish in the UV, \ie at 
large momenta as expected on $SO(6)$ R-symmetry grounds.

Altough the above integral cannot be expressed in terms of elementary
functions, it is not difficult to integrate the products of
hypergeometric functions by series. Moreover the
expressions drastically simplify when one goes on-shell \ie $p_i^2
= - m_i^2$ and amputates the external legs thus effectively
computing the irreducible vertex that enters the 3-body
scattering amplitude or, equivalently, the superglueball decay amplitude.

To this end observe that near the mass-shell where
\be
p_i^2L^2 =
- m(k_i)^2 L^2 + \epsilon_i
\ee
 \be a_-(p_i) \approx - k_i -
{\epsilon_i \over 4 (2 k_i +3)} \quad {\rm and} \quad a_+(p_i)
\approx 3 + k_i + {\epsilon_i \over 4 (2 k_i +3)} \ee
 This is true both for $\sigma$ and $\phi$.
Moreover,
at the mass-shell the hypergeometric functions truncate and become degree
$k$ (Gegenbauer or Jacobi) polynomials.

The mass poles of the connected 3-point function 
${\cal G}_{_{GPPZ}}(p_1,p_2, p_3)$ are exposed
by the $\Gamma$'s in front of the integral. Indeed \be
\Gamma(a_-(p_i)) \approx \Gamma\left(- k_i - {\epsilon_i \over 4 (2 k_i
+3)}\right) \approx {(-)^{k_i +1} 4 (2k_i + 3) \over k_i! \epsilon_i}
\ee while \be \Gamma(a_+(p_i)) \approx \Gamma\left(3 + k_i +
{\epsilon_i \over 4 (2 k_i +3)}\right) \approx (k_i + 2)! \ee 
The relevant integrals are of the form \bea &&\int_0^1 du (1-u)^2
u^{\Sigma n + 1} [2n_1 n_2 (1-u) - (n_1+n_2) u]
=  \\
&&{ 3! 2n_1 n_2 \over (\sum n + 2)(\sum n + 3)(\sum n + 4)(\sum n
+ 5)} - { 2!(n_1+n_2) \over (\sum n + 3)(\sum n + 4)(\sum n + 5)}
\nonumber \eea 
replacing the Pochhammer symbols in the hypergeometric functions
with their explicit expressions 
\be (-k_i)_{n_i}
= {(-)^{n_i} k_i! \over (k_i - n_i)!} \qquad (3 + k_i)_{n_i}
= {(k_i + n_i+2)!\over (k_i + 2)!} \ee and putting 
everything together yields \bea && A_{_{GPPZ}}
(k_1,k_2,k_3) = {(2\pi)^5\sqrt{3}\over N L}
\prod_{i=1}^3\sqrt{(k_i + 1)(k_i + 2)(2k_i + 3)} \times
\\
&&\sum_{n_i =0}^{k_i}\left\{ { 6 n_1 n_2 - (n_1 + n_2)(\sum n + 2)
\over (n_3 + 2)\prod_{s=2}^5(\sum n + s)}
\prod_{i=1}^3\left[{(-)^{k_i + n_i} \over n_i! (n_i+1 )!} {(k_i +
n_i+2)!\over (k_i - n_i)!}\right]\right\} \nonumber \eea
which has the expected mass dimension, $[{\rm m}]^1$ and
is correctly suppressed by the $1/N$ factor 
at large $N$. Comparison with field theory 
results seems prohibitive for the time being since it would require a 
non-perturbative understanding of the dynamics of ${\cal N}=1$ SYM which is 
well beyond reach. One may take the holographic result as a very 
precise prediction for the amplitude at strong 't Hooft coupling.

\section{Concluding remarks and speculations}
\label{fine}

We have studied three-point functions of scalar operators along two
holographic RG flows. For the CB flow we have considered inert scalars and 
derived a compact expression that is finite and behaves correctly in the UV.
In the GPPZ flow, though the initial situation was subtler because of the 
mixing of $\F$ with the metric, 
we have been able to extract the irreducible vertices of three 
superglueballs. Thanks to the operator identity
\be
 T_i{}^i =  \beta \co_\phi 
\ee
with $\beta= - \sqrt{3}$,
the very same modes couple to the trace of the stress tensor \cite{bfs1,wm}.
This should elucidate the evolution with scale according to the 
holographic renormalization group \cite{erd}.
Extending the analysis to the transverse modes of the stress tensor or of the 
(broken) currents is straightforward in principle since the relevant 
trilinear bulk 
couplings are completely fixed by gauge invariance. The expected form of 
these correlators should be
\bea
&& \< T_{ij}(p_1) \co_\sigma (p_2) \co_\sigma (p_3)\>_t 
= \kappa_{_T} 
 \delta(\sum_i p_i) \int dw \sqrt{G(w)} K^t_{ij,kl}(p_1,w)\times
\\
&&\times [ e^{-2A(w)}
(p_2^k p_3^l +p_2^l p_3^k) 
+ {4\over 3} \delta^{kl} M^2_\sigma(w)]  
K_\sigma(p_2,w) K_\sigma(p_3,w) + {\rm contact}
\nonumber
\eea
and
\bea
&& \< J_{i}(p_1) \co_\sigma (p_2) \co_{\tilde\sigma} (p_3)\>_t 
= \kappa_{_J} \delta(\sum_i p_i)
\int dw \sqrt{G(w)} K^t_{i,k}(p_1,w) \times
\nonumber\\
&& (p_2^k - p_3^k) 
K_\sigma(p_2,w) K_{\tilde\sigma}(p_3,w) + {\rm contact} 
\eea
As expected, the integrals diverge but the contact 
terms suggested by holographic renormalization cancel these divergences.
Similar considerations apply to three-point functions with more insertions 
of the stress tensor or of broken currents. As usual,
bulk gauge invariance translates into Ward
identities of the boundary quantum field theory. The
holographically computed one-point functions in the presence of sources
do satisfy these Ward identities including anomalies and this carries over to 
higher point functions.

It would be nice to address similar issues in fullfledged string solutions
that display confinement and logarithmic corrections to the pure AdS behavior 
in the UV \cite{mn,ks,kt}.
The results obtained by Krasnitz for the case of
two-point functions in the KT flow are very encouraging in this respect 
\cite{kras}.

As far as higher than three point correlation 
functions are concerned, if the miracle
that in AdS transforms exchange diagrams into contact ones does not take place
in non-trivial 
holographic RG flows then three-point functions will set the standard of a
new holographic tradition, at least for a 
while. 
 
\section*{Acknowledgements}

We would like to thank  S.~Kovacs, W.~M\"uck,
A.~Sagnotti, and Ya.~Stanev for useful 
discussions and especially M.~Berg and D.~Freedman for valuable
comments on the manuscript. While this work was being completed 
M.B. was visiting MIT as part of the INFN-MIT ``Bruno Rossi'' exchange program.
This work was supported in part by I.N.F.N., by the
EC programs HPRN-CT-2000-00122 and
HPRN-CT-2000-00148, by the INTAS contract
99-1-590, by the MURST-COFIN contract 2001-025492 and by the NATO
contract PST.CLG.978785.

\section*{Appendix: Conformal 3-point function in momentum space}

In order to have a glimpse of the result of the integral for
the holographic 3-point correlation function in the CB flow and
to eventually compare with not-so-standard field theory
results it is convenient to compute the Fourier transform of the
3-point function of scalar primary operators of dimension
$\Delta_i$ $i=1,2,3$. 
\be {\cal G}_{_{CFT}}(x_1,x_2,x_3) = \langle \co (x_1) \co
(x_2) \co (x_3) \rangle = \kappa_{_{CFT}} (x_{12}^2)^{-\Delta_{12}}
(x_{23}^2)^{-\Delta_{23}}(x_{31}^2)^{-\Delta_{31}} \ee where
$x_{ij}^2 = (x_i - x_j)^2$ and $\Delta_{ij}= \Delta_i + \Delta_j -
\Sigma$ with $\Sigma = (\Delta_1 + \Delta_2 + \Delta_3)/2$.
For simplicity we set the constant $\kappa_{_{CFT}} =1$ henceforth.
By definition \be {\cal G}_{_{CFT}}(p_1,p_2,p_3) = \int d^4x_1 d^4x_2 d^4x_3
e^{-i(p_1x_1+p_2x_2+p_3x_3)}
{\cal G}_{_{CFT}}(x_1,x_2,x_3)
\ee 
In $d$ dimensions, Feynman parametrization leads to the integral  \bea
&&{\cal G}_{_{CFT}}(p_1,p_2,p_3) = \delta(\sum_i p_i)
 {\pi^{(\Sigma+d)/2}
\Gamma(d/2 - \Delta_{23})  \Gamma(d - \Sigma)
\over 2^\Sigma \Gamma(d/2) \Gamma(\Delta_{12})  \Gamma(\Delta_{13})}\times 
\nonumber\\
&&\times\int_0^1 d\sigma F(\Sigma - {d\over 2}, \Delta_{23};{d\over 2};
1-\xi(\sigma)) {(1-\sigma)^{d/2 - \Delta_{13}-1} \sigma^{d/2 -
\Delta_{12}-1} \over [(1-\sigma)(p_3+p_2)^2 +  \sigma
p_2^2]^{d-\Sigma}} \nonumber 
\eea 
where \be \xi(\sigma) =
{(1-\sigma)\sigma p_3^2 \over (1-\sigma)(p_3+p_2)^2 +  \sigma
p_2^2} \ee We have checked that this expression correctly factorizes in the
extremal case $\Delta_1 = \Delta_2 + \Delta_3$.

Let us specialize to the case $\Delta_i=2$  in $d=4$, so that $\Sigma=3$
and $\Delta_{ij} = 1$. Putting 
\be 
\mu_{2}\equiv p_2^2/p_1^2 \qquad
\mu_{3}\equiv p_3^2/p_1^2 
\ee
one gets
\be 
{\cal G}_{_{CFT}}(p_1,p_2,p_3)
= {\pi^{7/2}\over 8 p_1^2} \delta(\sum_i p_i) 
A(\mu_{2},\mu_{3})
\ee
where
\be
A(\mu_{2},\mu_{3})\equiv \int_0^1 d\sigma { \log
[(1-\sigma)\sigma] - \log[\sigma \mu_{2} + (1-\sigma) \mu_{3}]
\over (1-\sigma)\sigma +  \sigma \mu_{2} + (1-\sigma)\mu_{3}} \ee
is strikingly reminiscent of the box integral $B(r,s)$ computed \eg in 
\cite{bkrs}. Indeed $A(\mu_{2},\mu_{3})$ is tightly related to  $B(r,s)$. 

It is easy to check the symmetry 
$A(\mu_{2},\mu_{3})= A(\mu_{3},\mu_{2})$. It is slightly more laborious to 
express it in terms of logs and dilogs. To this end it is convenient to
factorize the denominator as
\be \sigma(1-\sigma)+\mu_{2}\sigma
+(1-\sigma)\mu_{3}=-(\sigma-\sigma_{+})(\sigma-\sigma_{-}) \ee
where
\be
\sigma_{\pm} = {1\over 2}
[1+\mu_2 - \mu_3 \pm\sqrt{(1+\mu_2 - \mu_3)^2 + 4\mu_3}]
\ee
Repeatedly and carefully (since they are valid for $0\leq x\leq 1$)
using 
\be 
Li_2(x) + Li_2(1-x)=-\log(x)\log(1-x)+\frac{\pi^{2}}{6} 
\label{dil1}
\ee
and
\be  
Li_2(x) + Li_2\bl\frac{x-1}{x}\br=-\frac{1}{2}\log^{2}(x) 
\label{dil2} 
\ee 
where\footnote{In the notation of Abramowitz-Stegun \cite{as} (pag.1004) 
$Li_{2}(z)=f(1-z)$.}
\be 
Li_2(z) = - \int_0^1 {d\sigma \over
\sigma}\log(1-z\sigma) = \sum_{n=1}^\infty {z^n\over n^2}
\label{dilog} 
\ee  
one finally gets the desired result
\bea 
&&A(\mu_{2},\mu_{3}) = \frac{1}{\sqrt{\nu}}\bl
2Li_{2}\left( {1 + \mu_2 - \mu_3 - \sqrt{\nu}\over 2 \mu_2}\right) 
+
2Li_{2}\left( {1 + \mu_2 - \mu_3 - \sqrt{\nu}\over 2 \mu_2}\right)
\nonumber \\
&& - \log(\mu_2)\log(\mu_3) - 
\left[\log\left( {\mu_2 +\mu_3 - 1 - \sqrt{\nu}\over 2}\right)\right]^2
 \br  
\eea
where
\be
 \nu = 1 + \mu^2_2 + \mu_3^2 - 2\mu_1 - 2\mu_2 - 2 \mu_2 \mu_3
\ee
$A(\mu_2,\mu_3)$ is manifestly symmetric and real for $\nu\ge 0$.
Up to an overall constant, 
$A(\mu_2,\mu_3)$ coincides with $B(r,s)$ or better $g(x_1,x_2,x_3)$
in \cite{bkrs} after replacing $\mu_2$ and $\mu_3$ with $r$ and $s$, \ie 
$p_1\to x_{23}$,$p_2\to x_{31}$, $p_3\to x_{12}$,

\section*{Appendix B: Other useful formulae}

For the RG flows under consideration, homogeneous solution of the 
fluctuation equations may be expressed in terms of hypergeometric functions
 \be 
\phi_p(w,x) = w^\a(1-w)^\b F(a,b;c;w)\exp(ipx) 
\ee .
We assume that $z=0$ represents the deep IR interior (singularity) 
and $w=1-\epsilon$ the (regulated) UV boundary.  
In order to fix the overall normalization of the 
(regulated) bulk-to-boundary propagators $K_\epsilon(p,z)$ 
we need the analytic continuation of $F(a,b,c;w)$ to $w\approx 1$.

For $c=a+b$ \bea &&F(a,b;a+b;w)= - \ { \G(a+b) \over \G(a) \G(b)}
\sum_{n=0}^{\infty}{(a)_n (b)_n \over (n!)^2 } (1-w)^n \times
 \\
&&\times[\log(1-w)-2\psi(n+1)+\psi(a+n)+\psi(b+n)] \nonumber \eea

For $c=a+b-m$ \bea &&F(a,b;a+b-m;w)= {\G(m) \G(a+b-m) \over \G(a)
\G(b)} \sum_{n=0}^{m-1}{(a-m)_n (b-m)_n \over n! (1-m)_n} \times
\nonumber\\
&&\times (1-w)^{n-m} \ - \ {(-)^m \G(a+b-m) \over \G(a-m) \G(b-m)}
\sum_{n=0}^{\infty}{(a)_n (b)_n \over n! (n+m)!} (1-w)^n \times
 \\
&&\times[\log(1-w)-\psi(n+1)-\psi(n+m+1)+\psi(a+n)+\psi(b+n)]
\nonumber \eea 

For $c=a+b+m$ \bea &&F(a,b;a+b+m;w)= {\G(m)
\G(a+b+m) \over \G(a+m) \G(b+m)} \sum_{n=0}^{m-1}{(a)_n (b)_n
\over n! (1-m)_n} (1-w)^{n}
 \nonumber \\
&& - \ { (-)^m \G(a+b+m) \over \G(a) \G(b)}
\sum_{n=0}^{\infty}{(a+m)_n (b+m)_n \over n! (n+m)!} (1-w)^{n+m} \times
\\
&&\times[\log(1-w)-\psi(n+1)-\psi(n+m+1)+\psi(a+n+m)+\psi(b+n+m)]
\nonumber \eea

Correspondingly, the (regulated) bulk to boundary propagators read 

\bea &&K_\epsilon(p,w) = - {w^\a(1-w)^\b
\G(a)\G(b)\over\epsilon^\beta \log\epsilon \G(a+b)} F(a,b;a+b;w) 
\\
&&K_\epsilon(p,w) ={ w^\a(1-w)^{\b}\G(a)
\G(b)\over\epsilon^{\b}\G(m)\G(a+b-m)} F(a,b;a+b-m;w)
\\
&&K_\epsilon(p,w) ={ w^\a(1-w)^\b
\G(a+m)\G(b+m)\over\epsilon^{\b} \G(m)\G(a+b+m)} F(a,b;a+b+m;w) 
\eea

\end{document}